\documentstyle[prd,eqsecnum,aps]{revtex}
\tighten
\begin{document}
\draft
\title{Remarks on flavor-neutrino propagators and oscillation formulae}
\author{Kanji Fujii$^{*1}$, Chikage Habe$^{*1}$
 and Tetsuo Yabuki$^{*2}$
\\
{\it 
\begin{tabular}{l}
$^{*1}$ Department of Physics, Faculty of Science, \\
\hspace*{8mm}Hokkaido University, Sapporo 060-0810, Japan\\
$^{*2}$ Rakunou Gakuen University, Ebetsu 069-0836, Japan
\end{tabular}}
}
\date{\today}

\maketitle
\begin{abstract}
We examine the general structure of the formulae of neutrino oscillations proposed by Blasone and Vitiello(BV).
Reconstructing their formulae with the retarded propagators of the flavor neutrino fields for the case of many flavors, we can get easily the formulae which satisfy the suitable boundary conditions and are independent of arbitrary mass parameters $\{\mu_{\rho}\}$, as is obtained by BV for the case of two flavors.
In this two flavor case, our formulae reduce to those obtained by BV under $T$-invariance condition.
Furthermore, the reconstructed probabilities are shown to coincide with those derived with recourse to the mass Hilbert space ${\cal H}_{m}$ which is unitarily inequivalent to the flavor Hilbert space ${\cal H}_{f}$.
Such a situation is not found in the corresponding construction a la BV.
Then the new factors in the BV's formulae, which modify the usual oscillation formulae, are not the trace of the flavor Hilbert space construction, but come from Bogolyubov transformation among the operators of spin-$\frac{1}{2}$ neutrino with different masses.
\end{abstract}

\pacs{PACS number(s):14.60.Pq }

\section{Purpose and fundamental assumption}

     The field theoretical descriptions of neutrino oscillations have been examined from various viewpoints\cite{FT,GKL,BV1,FHY,BV2,BHV}.  
When we want to reformulate straightforwardly, in the framework of the field theory, the familiar quantum-mechanical derivation of the neutrino oscillation formula\cite{BP}, we encounter the problem how to define field-theoretically one (anti-)neutrino state with a definite flavor.   
Giunti et al.\cite{GKL} gave a negative answer to this problem on the basis of the observation that the Hilbert space of the weak eigenstates with definite flavors can be constructed approximately only in extremely relativistic case.\footnote{This assertion seems to be not so convincing, the reason of which will be explained in Appendix C.}   
    In Ref.\cite{BV1}, the authors asserted that the flavor (or weak) as well as the mass Hilbert spaces, ${\cal H}_{f}$ and ${\cal H}_{m}$, can be really constructed by employing the Bogolyubov transformation among creation and annihilation operators of the flavor and mass eigenstates of neutrinos. 
The unitary inequivalence between those two Hilbert spaces leads to a certain effect in the neutrino oscillation formulae, which is to be observed in the low-energy experiment.        

    In order to determine the coefficients appearing in the Bogolyubov transformation mentioned above, the masses of the electron- and muon-neutrinos (in the 2-flavor case) were taken in Ref.\cite{BV1} to be the mass eigenvalues $m_{1}$  and $m_{2}$, respectively.  
To this prescription, the present authors gave a criticism\cite{FHY}. 
Its essence lies in the point that the masses of flavor neutrinos are inherently arbitrary and such arbitrariness should not remain in any observed quantities; thus, it is unphysical that the oscillation formulae of neutrinos depend on the arbitrarily chosen mass parameters.

     In connection with this criticism\cite{FHY}, Blasone and Vitiello (hereafter abbreviated as BV)\cite{BV2} have remarked that there exist some quantities leading to the neutrino oscillation formulae, which satisfy the necessary boundary conditions and also are independent of the mass parameters of flavor neutrinos even when we start with the theory including such arbitrary parameters.

     The purpose of the present paper is to present clearly, on a general basis of the field theory, the logical feature of the remark given by BV in Ref.\cite{BV2}.  
The considerations developed in the following are based on the $<$Setup$>$: -- The relation of the flavor neutrino field operator  $\nu_{\rho}(x)$ to the neutrino field operator $\nu_{j}(x)$ is expressed as 
\begin{eqnarray}\label{n}
\nu_{\rho}(x)=\sum_{j=1}^{N_{f}}z^{1/2}_{\rho j}\nu_{j}(x),\mbox{\hspace*{5mm}}\rho=e,\mu,\tau,\cdots;
\end{eqnarray}
here $\nu_{j}(x)$ satisfies the free Dirac equation with a definite mass  $m_{j}$,
\begin{eqnarray}
            (\not\partial+ m_{j})\nu_{j}(x) = 0, 
\end{eqnarray}
and the matrix  $Z^{1/2}  = ( z^{1/2}_{\rho j}    )$ satisfies    $\sum_{j=1}^{N_{f}}z^{1/2}_{\rho j}z^{1/2\ast}_{\sigma j}=\delta_{\rho\sigma} ; N_{f} =\mbox{ the number of flavors}$.

The linear combinations (\ref{n}) are determined so as to diagonalize the mass term in the Lagrangian ( in other words, so as to diagonalize the pole part of the propagator matrix constructed from the flavor neutrino field operators\cite{KOW}; when only the repetitions of the bilinear mass-type interaction are taken into account, the unitarity of $Z^{1/2}$ is obtained\cite{FHY}.
If the CP- (or T-)invariance is required, we obtain the reality of $Z^{1/2}$\cite{KOW}.)

    Due to the above setup, we obtain the c-number property of the anticommutators 
\begin{eqnarray}\label{com}
      \{\nu_{\rho}(x),\nu_{\sigma}^{\dagger}(y) \} = \mbox{c-number},
\end{eqnarray}
and 
\begin{eqnarray}\label{ac}
      \{\nu_{\rho}(x),\nu_{\sigma}(y) \} = 0,
\end{eqnarray}
Further, the canonical commutation relations among the flavor neutrino field operators at an equal time are consistently obtained;
\begin{eqnarray}\label{comrel}
      \{\nu_{\rho}(\vec x,t),\nu_{\sigma}^{\dagger}(\vec y,t) \} = \delta_{\rho\sigma}\delta (\vec x-\vec y) ,\mbox{\hspace*{5mm}}  \{\nu_{\rho}(\vec x,t),\nu_{\sigma}(\vec y,t) \} = 0         
\end{eqnarray}
due to the unitarity of  $Z^{1/2}$ and due to (\ref{ac}) respectively.

     Therefore, from (\ref{n}) we see that 
\begin{eqnarray}
      \langle vac |\{\nu_{\rho}(x),\bar\nu_{\sigma}(y) \}|vac\rangle  
\end{eqnarray}
does not depend on the choice of the vacuum state when this state is equally normalized as  $\langle vac | vac\rangle   = 1$.
   In other words, the expectation value
\begin{eqnarray}
      {}_{f}\langle 0(T) |\{\nu_{\rho}(x),\bar\nu_{\sigma}(y) \}|0(T)\rangle  _{f},
\end{eqnarray}
where   $|0(T) \rangle  _{f}$   is the vacuum state (at an arbitrary time $T$) belonging to the flavor Hilbert space ${\cal H}_{f}$ specified by a set of the mass parameters $\{\mu_{e},\mu_{\mu}, \cdots \} = \{\mu_{\lambda}\}$, is equal to 
\begin{eqnarray}
      {}_{m}\langle 0 |\{\nu_{\rho}(x),\bar\nu_{\sigma}(y) \}|0\rangle  _{m},\mbox{\hspace*{1cm}}|0\rangle  _{m}\in{\cal H}_{m};
\end{eqnarray}
this equality holds irrespectively of both $\{\mu_{\lambda}\}$ and the time $T$, since the c-number in (\ref{com}) depends on $\{m_{j}\}$ but not on $\{\mu_{\lambda}\}$ due to the $<$Setup$>$.  
(As to the definitions of the vacuum states, see the next section.)

     In the following sections, we will explain that these facts described above provide a general field theoretical basis for understanding the implications included in the remark given in Ref.\cite{BV2}. 

     In Sec.II, we summarize the essence of Ref.\cite{BV2} after giving necessary definitions of notations and relations.  
In Sec.III we make clear, on the basis of the $<$Setup$>$, the general implications of the BV's remark\cite{BV2}.  
Sec. IV is devoted to summarizing discussions. In Appendices, relevant mathematical details are given.

\section{Reformulation of BV's work and related remark}
\subsection{Notations and definitions}
We summarize the notations and definitions of the related quantities in accordance with Ref.\cite{FHY}.

The relation (\ref{n}) between the flavor eigenfields, $\nu_{F}$, and the mass eigenfields, $\nu_{M}$, is expressed by the transformation as

\begin{eqnarray}
{\bf\nu}_{F}(x) &\equiv& \left(\begin{array}{c}
\nu_{e}(x)\\\nu_{\mu}(x)\\ \vdots
\end{array}\right)
=G^{-1}(x^{0})\left(\begin{array}{c}
\nu_{1}(x)\\\nu_{2}(x)\\ \vdots
\end{array}\right)G(x^{0})
=\left(\begin{array}{ccc}
z_{e1}^{1/2}&z_{e2}^{1/2}& \\
z_{\mu 1}^{1/2}&z_{\mu 2}^{1/2}& \\
&&\ddots
\end{array}\right)
\left(\begin{array}{c}
\nu_{1}(x)\\\nu_{2}(x)\\ \vdots
\end{array}\right)\nonumber\\
&\equiv&Z^{1/2}{\bf\nu}_{M}(x),
\end{eqnarray}
where
\begin{eqnarray}
Z^{1/2}=[z^{1/2}_{\rho j}],\mbox{\hspace*{1cm}}Z^{1/2}Z^{1/2\dagger}=I;
\end{eqnarray}
the concrete form of $G(x^{0})$ in the two flavor case is given by BV\cite{BV1}.
Let us expand the neutrino field in terms of helicity-momentum eigenfunctions as
\begin{eqnarray}
\nu_{a}(x)&=&\frac{1}{\sqrt{V}}\sum_{\vec k r}\{u_{a}(kr)\alpha_{a}(kr;t)e^{i\vec k\cdot\vec x}+v_{a}(kr)\beta^{\dagger}_{a}(kr;t)e^{-i\vec k\cdot\vec x}\}\nonumber\\
&=&\frac{1}{\sqrt{V}}\sum_{\vec k r}e^{i\vec k\cdot\vec x}\{u_{a}(kr)\alpha_{a}(kr;t)+v_{a}(-kr)\beta^{\dagger}_{a}(-kr;t)\},
\end{eqnarray}
where in the Kramers representation
\begin{eqnarray}\label{rholambda}
&&(i\not k +\mu_{a})u_{a}(kr)=0,(i\not k -\mu_{a})v_{a}(kr)=0,k_{0}=\sqrt{\vec k^{2}+\mu_{a}^{2}}\equiv \omega_{a}(k),\nonumber\\
&&u_{a}^{\ast}(kr)u_{b}(ks)=v_{a}^{\ast}(-kr)v_{b}(-ks)=\rho_{ab}(k)\delta_{rs},\nonumber\\
&&u_{a}^{\ast}(kr)v_{b}(-ks)=v_{a}^{\ast}(-kr)u_{b}(ks)=i\lambda_{ab}(k)\delta_{rs},\nonumber\\
&&\rho_{ab}(k)\equiv cos(\frac{\chi_{a}-\chi_{b}}{2}),\lambda_{ab}(k)\equiv sin(\frac{\chi_{a}-\chi_{b}}{2}), cot\chi_{a}=\frac{|\vec k|}{\mu_{a}}.
\end{eqnarray}
(For the case that $\nu_{a}(x)$ represents the mass eigenfield $\nu_{j}(x)$, we write $\mu_{j}$ as $m_{j}$.
Note that, for $a=\lambda(=e,\mu,\cdots)$, $\mu_{\lambda}$ is an arbitrarily fixed parameter\cite{FHY}.
As to the concrete forms of $u_{a}(kr)$ and $v_{a}(kr)$, see App.C.)
Here we use the notations
\begin{eqnarray}
{\bf\alpha}_{F}(kr;t) &\equiv& \left(\begin{array}{c}
\alpha_{e}(kr;t)\\ \alpha_{\mu}(kr;t)\\ \vdots
\end{array}\right),
{\bf\beta}_{F}(-kr;t) \equiv \left(\begin{array}{c}
\beta_{e}(-kr;t)\\ \beta_{\mu}(-kr;t)\\ \vdots
\end{array}\right),\nonumber\\
{\bf\alpha}_{M}(kr;t) &\equiv& \left(\begin{array}{c}
\alpha_{1}(kr;t)\\ \alpha_{2}(kr;t)\\ \vdots
\end{array}\right),
{\bf\beta}_{M}(-kr;t) \equiv \left(\begin{array}{c}
\beta_{1}(-kr;t)\\ \beta_{2}(-kr;t)\\ \vdots
\end{array}\right).
\end{eqnarray}
We have
\begin{eqnarray}\label{calK}
\left(\begin{array}{c}
\alpha_{F}(kr;t)\\\beta_{F}^{\dagger}(-kr;t)
\end{array}\right)
=K^{-1}(t)\left(\begin{array}{c}
\alpha_{M}(kr;t)\\\beta_{M}^{\dagger}(-kr;t)
\end{array}\right)K(t)
={\cal K}(k)\left(\begin{array}{c}
\alpha_{M}(kr;t)\\\beta_{M}^{\dagger}(-kr;t)
\end{array}\right),
\end{eqnarray}
with
\begin{eqnarray}
{\cal K}(k)&=&\left(\begin{array}{cc}
P(k)&i\Lambda(k)\\i\Lambda(k)&P(k)
\end{array}\right), \mbox{\hspace*{5mm}}  {\cal K}(k){\cal K}^{\dagger}(k)=I,\nonumber\\
&&P(k)=[P(k)_{\rho j}]=[z^{1/2}_{\rho j}\rho_{\rho j}(k)], \Lambda(k)=[z^{1/2}_{\rho j}\lambda_{\rho j}(k)].
\end{eqnarray}
${\cal K}$ is independent of the time $t$, but depends, of course, on $\{\mu_{\lambda}\}$ and $[z^{1/2}]_{\rho j}$ other than $|\vec k|$, and we dropped such dependence for simplicity.
The relation between the vacuum states, $|0(t)\rangle  _{f}\in{\cal H}_{f}$ and $|0\rangle  _{m}\in{\cal H}_{m}$, is expressed with this $K(t)$ as
\begin{eqnarray}
|0(t)\rangle  _{f}=K(t)^{-1}|0\rangle  _{m}.
\end{eqnarray}
Here, these vacua are defined for ${}^{\forall}\vec k$ and $r$ as
\begin{eqnarray}
\alpha_{F}(kr;t)|0(t)\rangle  _{f}=\beta_{F}(kr;t)|0(t)\rangle  _{f}=0, \mbox{\hspace*{1cm}}\alpha_{M}(kr;t)|0\rangle  _{m}=\beta_{M}(kr;t)|0\rangle  _{m}=0
\end{eqnarray}
with the normalization $_{f}\langle 0(t)|0(t)\rangle  _{f}={}_{m}\langle 0|0\rangle  _{m}=1$.

>From (\ref{calK}), we obtain the relations connecting the creation and annihilation operators with different times, expressed as
\begin{eqnarray}\label{DefOfW}
\left(\begin{array}{c}
\alpha_{F}(kr;0)\\\beta_{F}^{\dagger}(-kr;0)
\end{array}\right)
={\cal K}(k)\left(\begin{array}{c}
\alpha_{M}(kr;0)\\\beta_{M}^{\dagger}(-kr;0)
\end{array}\right)
=W(k;t)\left(\begin{array}{c}
\alpha_{F}(kr;t)\\\beta_{F}^{\dagger}(-kr;t)
\end{array}\right)
\end{eqnarray}
with
\begin{eqnarray}\label{W}
W(k;t)&\equiv&{\cal K}(k)\Phi(k;t){\cal K}^{\dagger}(k)
=\left(\begin{array}{cc}
P\phi P^{\dagger}+\Lambda\phi^{\ast}\Lambda^{\dagger}&i(-P\phi \Lambda^{\dagger}+\Lambda\phi^{\ast}P^{\dagger})\\
i(\Lambda\phi P^{\dagger}-P\phi^{\ast}\Lambda^{\dagger})&\Lambda\phi \Lambda^{\dagger}+P\phi^{\ast}P^{\dagger})
\end{array}\right),
\nonumber\\
\Phi(k;t)&\equiv&\left(\begin{array}{cc}
\phi(t)&0\\0&\phi^{\dagger}(t)
\end{array}\right),\mbox{\hspace*{5mm}}
\phi(t)=\left(\begin{array}{ccc}
e^{i\omega_{1} t}&&0\\&e^{i\omega_{2} t}&\\0&&\ddots
\end{array}\right).
\end{eqnarray}
Therefore we obtain
\begin{eqnarray}
K(t)\left(\begin{array}{c}
\alpha_{F}(kr;0)\\\beta_{F}^{\dagger}(-kr;0)
\end{array}\right)K^{-1}(t)
=W(k;t)\left(\begin{array}{c}
\alpha_{M}(kr;t)\\\beta_{M}^{\dagger}(-kr;t)
\end{array}\right),
\end{eqnarray}
or
\begin{eqnarray}
K(0)\left(\begin{array}{c}\alpha_{F}(kr;t)\\\beta_{F}^{\dagger}(-kr;t)\end{array}\right)K^{-1}(0)
=W^{\dagger}(k;t)\left(\begin{array}{c}\alpha_{M}(kr;0)\\\beta_{M}^{\dagger}(-kr;0)\end{array}\right).
\end{eqnarray}
We write the matrix elements of $W(k;t)$ as
\begin{eqnarray}
W(k;t)=\left(\begin{array}{cc}W_{\rho\sigma}(k;t)&W_{\rho\bar\sigma}(k;t)\\W_{\bar\rho\sigma}(k;t)&W_{\bar\rho\bar\sigma}(k;t)\end{array}\right).
\end{eqnarray}
>From (\ref{W}), we have
\begin{eqnarray}\label{Welement}
W_{\rho\sigma}(k;t)^{\ast}=W_{\bar\sigma\bar\rho}(k;t),\mbox{\hspace*{1cm}}
W_{\rho\bar\sigma}(k;t)^{\ast}=W_{\sigma\bar\rho}(k;t),\mbox{\hspace*{1cm}}
W_{\bar\rho\sigma}(k;t)^{\ast}=W_{\bar\sigma\rho}(k;t).
\end{eqnarray}
In addition to the unitarity of $W(k;t)$
\begin{eqnarray}
W(k;t)W^{\dagger}(k;t)=I
\end{eqnarray}
due to the unitarity of $Z^{1/2}$, we have
\begin{eqnarray}
W(k;T-t)W^{\dagger}(k;T)=W(k;-t)=W^{\dagger}(k;t).
\end{eqnarray}

\subsection{BV's results}
Let us review briefly the main contents of BV's paper\cite{BV2}.
For the two flavor case, we consider that an initial electron neutrino evolves(oscillates) in time with the two relevant propagators(for $t\ge 0$)
\begin{eqnarray}\label{propee}
i {G}^{>}_{ee}({\vec x},t;{\vec y},0) =    {}_{f}\langle
{0}(0)|\nu_{e}({\vec x},t) \;   \bar{\nu}_{e}({\vec y},0)
|{0}(0)\rangle_{f}, 
\\ \label{propem} 
i {G}^{>}_{\mu e} ({\vec x},t;{\vec y},0) =    {}_{f}\langle {
0}(0)|    \nu_{\mu}({\vec x},t) \; \bar{\nu}_{e}({\vec y},0)   |{
0}(0)\rangle_{f}.
\end{eqnarray}  
By employing their Fourier components
\begin{eqnarray}
iG^{>}_{\rho\sigma}(k;t)&\equiv&\frac{1}{V}\int d\vec x \int d\vec y iG^{>}_{\rho\sigma}(\vec x,t;\vec y,0)e^{-i\vec k(\vec x-\vec y)},\mbox{\hspace*{5mm}}\rho(\mbox{and } \sigma)=e, \mu,
\end{eqnarray}  
we define
\begin{eqnarray} \label{pee1}
{\widetilde {\cal P}}^r_{ee} ({k};t)   &\equiv& i \,u_{e}^{\dagger}(kr)\,   {G}^>_{ee}({k};t)\,\gamma^0 u_{e}(kr) =    \left\{{\alpha}_{e}(kr;t),{\alpha}^{\dagger}_{e}(kr;0) \right\},   \\ \nonumber  \\ \label{pee2}   
{\widetilde {\cal P}}^r_{\bar{e}e} ({k};t)&\equiv&i \,v_{e}^{\dagger}(-kr)\,   { G}^>_{ee}({k};t)\,\gamma^0 u_{e}(kr)=   \left\{{\beta}^{\dagger}_{e}(-kr;t), { \alpha}^{\dagger}_{e}(kr;0) \right\},   \\ \nonumber   \\ \label{pee3}   
{\widetilde {\cal P}}^r_{\mu e}({k};t)&\equiv& i \,u_{\mu}^{\dagger}(kr)\, {G}^>_{\mu e}({k};t)\,\gamma^0 u_{e}(kr) = \left\{{ \alpha}_{\mu}(kr;t),  { \alpha}^{\dagger}_{e}(kr;0) \right\},    \\ \nonumber   \\ \label{pee4}   
{\widetilde {\cal P}}^r_{\bar{\mu} e}({k};t)&\equiv&   i \,v_{\mu}^{\dagger}(-kr)\,   { G}^>_{\mu e}({k};t)\,\gamma^0 u_{e}(kr) =    \left\{{ \beta}^{\dagger}_{\mu}(-kr;t), { \alpha}^{\dagger}_{e}(kr;0) \right\}.   
\end{eqnarray}  
Then the quantities defined by
\begin{eqnarray} \label{enumber1}  
P_{\nu_e\rightarrow\nu_e}({k};t)&\equiv&|{\widetilde {\cal P}}^r_{ee} ({ k};t)|^{2}+|{\widetilde {\cal P}}^r_{\bar{e}e} ({k};t)|^{2}, \nonumber\\
P_{\nu_e\rightarrow\nu_\mu}({k};t)&\equiv&|{\widetilde {\cal P}}^r_{\mu e}({  k};t)|^{2}+|{\widetilde {\cal P}}^r_{\bar{\mu} e}({k};t)|^{2}
\end{eqnarray}  
are seen to be interpreted as the observable oscillation probabilities in the sense that these quantities satisfy the necessary boundary conditions as
\begin{eqnarray} \label{enumber2}  
P_{\nu_e\rightarrow\nu_{\rho}}({k};t=0)=\delta_{e\rho}, \mbox{\hspace*{5mm}}\sum_{\rho}P_{\nu_e\rightarrow\nu_{\rho}}({k};t)=1,
\end{eqnarray}  
and also are shown to be "$\mu_{\lambda}$-independent".
Therefore, the special choice of the mass parameters in Ref.\cite{BV2}, $\mu_{e}=m_{1}$ and $\mu_{\mu}=m_{2}$, is justified and free from the criticism of Ref.\cite{FHY}.

The resultant formulae of the probabilities are
\begin{eqnarray} \label{BVprob}  
P_{\nu_e\rightarrow\nu_e}({k};t)
&=& 1 - \sin^{2}( 2 \theta)\left[ \rho_{12}^{2}(k) \;    \sin^{2} \left( \frac{\omega_{2}(k) -\omega_{1}(k)}{2} t \right)   +\lambda_{12}^{2}(k) \;   \sin^{2} \left(\frac{\omega_{2}(k) + \omega_{1}(k)}{2} t \right) \right] \, , \nonumber\\
P_{\nu_e\rightarrow\nu_\mu}({k};t)&=& 1 -  P_{\nu_e\rightarrow\nu_e}({k};t).
\end{eqnarray}  
($\theta$ is the mixing angle in the two flavor case; see Eq.(\ref{Zfor2}).)
In the framework of Ref.\cite{BV2}, the new factors $\rho_{12}^{2}=1-\lambda_{12}^{2}$ appearing in the above oscillation formulae are thought to be a result of the unitary inequivalence between ${\cal H}_{f}$ and ${\cal H}_{m}$.

It is pointed out further by BV\cite{BV2} that the quantities in Eq.(\ref{BVprob}) coincide with the expectation values of the charge operators
\begin{eqnarray} 
Q_{\sigma}(t=0)\equiv
\sum_{\vec k, r}(\alpha_{\sigma}^{\dagger}(kr;0)\alpha_{\sigma}(kr;0) -
\beta^{\dagger}_{\sigma}(-kr;0)\beta_{\sigma}(-kr;0))\, , \sigma=e, \mu,
\end{eqnarray}
on the electron neutrino state at a time $t$
\begin{eqnarray}
|\nu_{e}(kr;t)\rangle  \equiv\alpha^{\dagger}_{e}(kr;t) |0(t)\rangle  _{f};
\end{eqnarray}
we have 
\begin{eqnarray} \label{charge1} 
&&\langle \nu_e(kr;t)|Q_\sigma(0)| \nu_e(kr;t)\rangle\, = \, 
\left|\left\{\alpha_{\sigma}(kr;0), \alpha^{\dag}_{e}(kr;t) 
\right\}\right|^{2}  
\;+ \;\left|\left\{\beta_{\sigma}^{\dag}(-kr;0),
\alpha^{\dag}_{e}(kr;t) \right\}\right|^{2}=P_{\nu_e\rightarrow\nu_{\sigma}}({k};t)
\, , 
\\[2mm] \label{charge2} 
&&\;_{f}\langle 0(t)|Q_\sigma(0)| 0(t)\rangle_{f}\, = \, 0 \,
\quad, \quad \langle \nu_e(kr;t)|\left(Q_e(0)\;+\;Q_{\mu}(0)\right)| \nu_e(kr;t)\rangle 
\,= \,1 \, .
\end{eqnarray}  

Those consequences of Ref.\cite{BV2} summarized above are confirmed by straightforwad calculations with the use of the concrete form of $W(k;t)$ in the two flavor case.
It seems, however, necessary for us to make clear a simple(or general) reason why the above consequances are obtained.
With this aim, we first rewrite the above consequances in the many flavor case, which is given in the next subsection, and will consider in Sec.III the structure of the retarded propagators of flavor neutrino fields.

Before entering the next subsection, it may be worthwhile to give a remark on the quantities appearing in (\ref{pee1})-(\ref{pee4}).
The definitions of $\tilde{\cal P}^{r}_{\rho e}(k;t)$ and $\tilde{\cal P}^{r}_{\bar\rho e}(k;t)$ employed by BV\cite{BV2} seems to be somewhat misleading.
These quantities are introduced only for convenience, and should not be understood as representing such as neutrino-antineutrino transitions to occur in neutrino oscillation process.
It is helpful for us to note that each of $\{\alpha_{\rho}(kr;t),\alpha^{\dagger}_{e}(kr;0)\}$ and $\{\beta^{\dagger}_{\rho}(-kr;t),\alpha^{\dagger}_{e}(kr;0)\}$ has only a nonvanishing term proportional to $\{\alpha_{\rho}(kr;0),\alpha^{\dagger}_{\sigma}(kr;0)\}$.
(Concretely, see (\ref{acW}))
Furhther we note that we obtain
\begin{eqnarray}
i {\bar G}^{>}_{\rho\sigma}({\vec x},t;{\vec y},0) \equiv   {}_{f}\langle
{0}(0)|\nu_{\rho}({\vec x},t) \;   {\nu}_{\sigma}({\vec y},0)
|{0}(0)\rangle_{f}=0.
\end{eqnarray}
due to $\{\alpha_{\rho}(qs;t),\beta^{\dagger}_{\sigma}(-kr;0)\}=\{\beta_{\rho}^{\dagger}(-qs;t),\beta^{\dagger}_{\sigma}(-kr;0)\}=0$ obtained from (\ref{ac}) or (\ref{DefOfW}).

\subsection{Rewriting  BV's formulae in many flavor case}
In order to study general structures of the BV's results, let us define the quantity in the many flavor case
\begin{eqnarray}\label{G>}
i{\cal G}^{>}_{\rho\sigma}(\vec x,t;\vec y,0)\equiv\theta(t){}_{f}\langle 0(0)|\nu_{\rho}({\vec x},t) \;   \bar{\nu}_{\sigma}({\vec y},0) |0(0)\rangle_{f}, \mbox{\hspace*{5mm}}\rho(\mbox{and }  \sigma)=e, \mu, \tau, \cdots.
\end{eqnarray}
($i{\cal G}^{>}_{\rho\sigma}(\vec x,0;\vec y,0)\equiv lim_{t\rightarrow +0}i{\cal G}^{>}_{\rho\sigma}(\vec x,t;\vec y,0)$.)
One can extract the component with the momentum $\vec k$ from this quantity as
\begin{eqnarray}
i{\cal G}^{>}_{\rho\sigma}(k;t)&\equiv&\frac{1}{V}\int d\vec x \int d\vec y i{\cal G}^{>}_{\rho\sigma}(\vec x,t;\vec y,0)e^{-i\vec k(\vec x-\vec y)}\nonumber \\
&=&\theta(t){}_{f}\langle 0(0)|\sum_{r}\left(\left\{\alpha_{\rho}(kr;t),\alpha^{\dagger}_{\sigma}(kr;0)\right\}u_{\rho}(kr){\bar u}_{\sigma}(kr)+\left\{\beta_{\rho}^{\dagger}(-kr;t),\alpha^{\dagger}_{\sigma}(kr;0)\right\}v_{\rho}(-kr){\bar u}_{\sigma}(kr)\right) |0(0)\rangle_{f}.
\end{eqnarray}
>From (\ref{DefOfW}), we obtain
\begin{eqnarray}\label{acW}
\left\{\alpha_{\rho}(kr;t),\alpha^{\dagger}_{\sigma}(kr;0)\right\}&=&\sum_{\kappa}\left\{[W^{\dagger}_{\rho\kappa}(k;t)\alpha_{\kappa}(kr;0)+W^{\dagger}_{\rho\bar \kappa}(k;t)\beta^{\dagger}_{\kappa}(-kr;0)]\mbox{\hspace*{2mm}},\alpha^{\dagger}_{\sigma}(kr;0)\right\}\nonumber\\
&=&W^{\dagger}_{\rho\sigma}(k;t)=W_{\sigma\rho}(k;t)^{\ast},\nonumber \\
\left\{\beta^{\dagger}_{\rho}(-kr;t),\alpha^{\dagger}_{\sigma}(kr;0)\right\}&=&\sum_{\kappa}\left\{[W^{\dagger}_{\bar\rho\kappa}(k;t)\alpha_{\kappa}(kr;0)+W^{\dagger}_{\bar\rho\bar \kappa}(k;t)\beta^{\dagger}_{\kappa}(-kr;0)]\mbox{\hspace*{2mm}},\alpha^{\dagger}_{\sigma}(kr;0)\right\}\nonumber\\
&=&W^{\dagger}_{\bar\rho\sigma}(k;t)=W_{\sigma\bar\rho}(k;t)^{\ast}.
\end{eqnarray}
Thus ${\cal G}^{>}_{\rho\sigma}(k;t)$ is given by
\begin{eqnarray}
i{\cal G}^{>}_{\rho\sigma}(k;t)=\theta(t)\sum_{r}\left(W_{\sigma\rho}(k;t)^{\ast}u_{\rho}(kr){\bar u}_{\sigma}(kr)+W_{\sigma\bar\rho}(k;t)^{\ast}v_{\rho}(-kr){\bar u}_{\sigma}(kr)\right).
\end{eqnarray}
With this quantity, we can define for $t\ge 0$
\begin{eqnarray}\label{P>}
P^{>}_{\nu_{\sigma}\rightarrow\nu_{\rho}}(k;t)&\equiv&\frac{1}{2}Tr[{\cal G}^{>}_{\rho\sigma}(k;t){\cal G}^{>\dagger}_{\rho\sigma}(k;t)]\nonumber\\
&=&|\left\{\alpha_{\rho}(kr;t),\alpha^{\dagger}_{\sigma}(kr;0)\right\}|^{2}+|\left\{\beta_{\rho}^{\dagger}(-kr;t),\alpha^{\dagger}_{\sigma}(kr;0)\right\}|^{2}\nonumber\\
&=&|W_{\sigma\rho}(k;t)^{\ast}|^{2}+|W_{\sigma\bar\rho}(k;t)^{\ast}|^{2},
\end{eqnarray}
which is equal to Eq.(\ref{BVprob}) for the two flavor case.
(Here, $Tr$ means to take the trace with respect to the indices of Dirac spinors.)
Along the same line as BV's which has been described in the previous subsection, let us call this quantity the pobability, since it satisfies automatically the normalization as
\begin{eqnarray}\label{norm}
\sum_{\rho}P^{>}_{\nu_{\sigma}\rightarrow\nu_{\rho}}(k;t)&=&1 
\end{eqnarray}
due to the unitarity of $W(k;t)$(or $Z^{1/2}$), and the boundary conditions as
\begin{eqnarray}\label{bound}
P^{>}_{\nu_{\sigma}\rightarrow\nu_{\rho}}(k;t=0)&=&\delta_{\rho\sigma}
\end{eqnarray}
due to the canonical communication relation at an equal time or the property of $W(k;t=0)=I$.

We cannot see straightforwardly the R.H.S. of (\ref{P>}) are independent of $\{\mu_{\lambda}\}$.
Such independency in the many flavor case is shown under the reality condition on $Z^{1/2}=[z^{1/2}_{\rho j}]$.
This condition is implicitly used in Ref.\cite{FHY}, where the 2-flavor case is examined; in this case

\begin{eqnarray}\label{Zfor2}
Z^{1/2}=\left(
\begin{array}{cc}
cos\theta & sin\theta \\
-sin\theta & cos\theta 
\end{array}
\right).
\end{eqnarray}
Under the condition of real $Z^{1/2}$, $P^{>}_{\nu_{\sigma}\rightarrow\nu_{\rho}}(k;t)$ is also equal to the expectation value of the number operator,
\begin{eqnarray}\label{DefOfN}
\langle N_{\sigma};kr;t\rangle  _{\rho-f}\equiv {}_{f}\langle 0(t)|\alpha_{\rho}(kr;t)N_{\sigma}(t=0)\alpha_{\rho}^{\dagger}(kr;t)|0(t)\rangle  _{f}.
\end{eqnarray}
Hereafter we use the notation $N_{\sigma}(t)$ instead of $Q_{\sigma}(t)$.

Detailed proofs of the $\{\mu_{\lambda}\}$-independence of Eq.(\ref{P>}) and the equality of the expectation value of the number operator to $P^{>}_{\nu_{\sigma}\rightarrow\nu_{\rho}}(k;t)$ are given in Appendices A and B.
\subsection{Corresponding propagator on ${\cal H}_{m}$}
It may be useful to note that the corresponding propagator defined on ${\cal H}_{m}$ does not have the same properties as $i{\cal G}^{>}_{\rho\sigma}(\vec x,t;\vec y,0)$.

The propagator which is constructed on ${\cal H}_{m}$ corresponding to (\ref{G>}) may be
\begin{eqnarray}
iS_{\rho\sigma}^{>}(\vec x,x^{0};\vec y,y^{0})&\equiv&\theta(x^{0}-y^{0}){}_{m}\langle 0|\nu_{\rho}(x)\bar \nu_{\sigma}(y)|0\rangle  _{m}\nonumber\\
&=&\sum_{j}z^{1/2}_{\rho j}z^{1/2\ast}_{\sigma j}\theta(x^{0}-y^{0}){}_{m}\langle 0|\left(\nu_{j}(x) \bar\nu_{j}(y)\right)|0\rangle  _{m},
\end{eqnarray}
which is a part of the Feynmann propagator, $iS_{F\rho\sigma}(x-y)=_{m}\langle 0|T\left(\nu_{\rho}(x)\bar \nu_{\sigma}(y)\right)|0\rangle  _{m}$.
By employing the quantity, in the same way as the case of the previous section,
\begin{eqnarray}
iS_{\rho\sigma}^{>}(k;t)&\equiv&
\frac{1}{V}\int d\vec x \int d\vec y iS^{>}_{\rho\sigma}(\vec x,t;\vec y,0)e^{-i\vec k(\vec x-\vec y)}
\end{eqnarray}
with $iS_{\rho\sigma}^{>}(k;t=0)\equiv lim_{t\rightarrow +0} iS_{\rho\sigma}^{>}(k;t)$, we obtain for $t\ge 0$
\begin{eqnarray}\label{Pi>}
\Pi^{>}_{\nu_{\sigma}\rightarrow\nu_{\rho}}(k;t)&\equiv&\frac{1}{2}Tr |S_{\rho\sigma}^{>}(k;t)S_{\rho\sigma}^{>\dagger}(k;t)|\nonumber\\
&=&\sum_{i,j}z^{1/2}_{\rho j}z^{1/2\ast}_{\sigma j}z^{1/2\ast}_{\rho i}z^{1/2}_{\sigma i}\rho^{2}_{ji}e^{i(\omega_{i}-\omega_{j})t}.
\end{eqnarray}
Note that this formula does not include the term proportional to $e^{\pm i(\omega_{i}+\omega_{j})t}$.
Although $\sum_{\rho}\Pi^{>}_{\nu_{e}\rightarrow\nu_{\rho}}(k;t)=1$ is satisfied, 
$\Pi^{>}_{\nu_{e}\rightarrow\nu_{\rho}}(k;t)$ does not satisfy the initial conditions as is easily seen in the two flavor case\footnote{Similar discussions can be seen in Ref.\cite{BHV}.}
\begin{eqnarray}
\Pi^{>}_{\nu_{e}\rightarrow\nu_{e}}(k;t)&\stackrel{t\rightarrow +0}{\longrightarrow}&1-\frac{1}{2}sin^{2}(2\theta)\lambda_{12}^{2}\neq 1,   \nonumber\\
\Pi^{>}_{\nu_{e}\rightarrow\nu_{\mu}}(k;t)&\stackrel{t\rightarrow +0}{\longrightarrow}&\frac{1}{2}sin^{2}(2\theta)\lambda_{12}^{2}\neq 0,   
\end{eqnarray}
and then is different from $P^{>}_{\nu_{e}\rightarrow\nu_{\rho}}(k;t)$ given in the subsection C.

One may say that this difference means the necessity of the flavor Hilbert space, ${\cal H}_{f}$.
We will examine in the next section whether it is true or not.

\section{Retarded kernel and amplitude}
Let us consider two types of the retarded propagators defined on ${\cal H}_{f}$ and ${\cal H}_{m}$ respectively as
\begin{eqnarray}
i{\cal G}^{(ret)}_{\rho\sigma}(\vec x,t;\vec y,0)&\equiv&\theta(t){}_{f}\langle 0(0)|\{\nu_{\rho}(\vec x,t), \bar \nu_{\sigma}(\vec y,0)\}|0(0)\rangle  _{f},\\
i{\cal S}^{(ret)}_{\rho\sigma}(\vec x,t;\vec y,0)&\equiv&\theta(t){}_{m}\langle 0|\{\nu_{\rho}(\vec x,t), \bar \nu_{\sigma}(\vec y,0)\}|0\rangle  _{m},
\end{eqnarray}
where these quantities at the time $t=0$ are defined by
\begin{eqnarray}
lim_{t\rightarrow +0}i{\cal G}^{(ret)}_{\rho\sigma}(\vec x,t;\vec y,0), \mbox{\hspace*{5mm}}lim_{t\rightarrow +0}i{\cal S}^{(ret)}_{\rho\sigma}(\vec x,t;\vec y,0).
\end{eqnarray}
Due to the c-number property of the anticommutator, the $\vec k$-components of these quantities are equal to each other
\footnote{
 We obtain not only for $T=0$ but also for an arbitrary time T 
\begin{eqnarray}
   _{f}\langle  0(T) |\{  \nu_{\rho}(\vec x,t), \bar \nu_{\sigma} (\vec y,0)\}  |0(T)\rangle  _{f} = _{m}\langle 0 |\{  \nu_{\rho}(\vec x,t), \bar \nu_{\sigma} (\vec y,0)\} | 0\rangle  _{m},\nonumber
\end{eqnarray}
where the vacuum states  $|0(T)\rangle  _{f}$  and $|0\rangle  _{m}$   are equally normalized.
Thus, the quantity
\begin{eqnarray}
      {\cal G}^{(ret)}_{\sigma\rho}(\vec x,t;\vec y,0;T) \equiv   _{f}\langle  0(T) |\{  \nu_{\rho}(\vec x,t), \bar \nu_{\sigma} (\vec y,0)\}  |0(T)\rangle  _{f}\nonumber
\end{eqnarray}
does not depend on $T$.
}
 as
\begin{eqnarray}
ig^{(ret)}_{\rho\sigma}(\vec k,t)= is^{(ret)}_{\rho\sigma}(\vec k,t);
\end{eqnarray}
therefore, one obtains the oscillation formula, which is independent of $\{\mu_{\lambda}\}$, as will be seen concretely in the following.

\subsection{The case of $i{\cal G}^{(ret)}_{\rho\sigma}(\vec x,t)$}
The $\vec k$-components of $i{\cal G}^{(ret)}_{\rho\sigma}(\vec x-\vec y,t)$ defined as
\begin{eqnarray}
ig^{(ret)}_{\rho\sigma}(k;t)\equiv \frac{1}{V}\int d\vec x\int d\vec y i{\cal G}^{(ret)}_{\rho\sigma}(\vec x,t;\vec y,0)e^{-i\vec k\vec x}e^{i\vec k\vec y}
\end{eqnarray}
become
\begin{eqnarray}
 ig^{(ret)}_{\rho\sigma}(k;t)&=&\theta(t)\sum_{r}\left(\left\{\alpha_{\rho}(kr;t),\alpha^{\dagger}_{\sigma}(kr;0)\right\}u_{\rho}(kr){\bar u}_{\sigma}(kr)+\left\{\beta_{\rho}^{\dagger}(-kr;t),\alpha_{\sigma}^{\dagger}(kr;0)\right\}v_{\rho}(-kr){\bar u}_{\sigma}(kr)\right.\nonumber\\
&&\left.+\left\{\alpha_{\rho}(kr;t),\beta_{\sigma}(-kr;0)\right\}u_{\rho}(kr){\bar v}_{\sigma}(-kr)+\left\{\beta_{\rho}^{\dagger}(-kr;t),\beta_{\sigma}(-kr;0)\right\}v_{\rho}(-kr){\bar v}_{\sigma}(-kr)\right).
\end{eqnarray}
With this quantity, we can define
\begin{eqnarray}
P^{(ret)}_{\nu_{\sigma}\rightarrow\nu_{\rho}}(k;t)\equiv\frac{1}{4}Tr [g^{(ret)}_{\rho\sigma}(k;t)g^{(ret)\dagger}_{\rho\sigma}(k;t)],
\end{eqnarray}
and call it the probability on the basis of its properties, similar to (\ref{norm}) and (\ref{bound}), as explained below.
For $t\geq 0$, we obtain
\begin{eqnarray}\label{Pret}
P^{(ret)}_{\nu_{\sigma}\rightarrow\nu_{\rho}}(k;t)&=&\frac{1}{4}\sum_{r}\left(|\left\{\alpha_{\rho}(kr;t),\alpha^{\dagger}_{\sigma}(kr;0)\right\}|^{2}+|\left\{\beta_{\rho}^{\dagger}(-kr;t),\alpha_{\sigma}^{\dagger}(kr;0)\right\}|^{2}\right.\nonumber\\
& &\left.+|\left\{\alpha_{\rho}(kr;t),\beta_{\sigma}(-kr;0)\right\}|^{2}+|\left\{\beta_{\rho}^{\dagger}(-kr;t),\beta_{\sigma}(-kr;0)\right\}|^{2}\right)\nonumber\\
&=&\frac{1}{2}\left(|W_{\sigma\rho}(k;t)|^{2}+|W_{\sigma\bar\rho}(k;t)|^{2}+|W_{\bar\sigma\rho}(k;t)|^{2}+|W_{\bar\sigma\bar\rho}(k;t)|^{2}\right),
\end{eqnarray}
due to Eq.(\ref{DefOfW})(or due to (\ref{a5}) and (\ref{a6})).
Clearly we can confirm the following properties of this quantity:
\begin{enumerate}
\renewcommand{\labelenumi}{(\arabic{enumi})}
\item Fundamentally due to the canonical commutation relation (\ref{comrel}),
\begin{eqnarray}\label{BC1}
P^{(ret)}_{\nu_{\sigma}\rightarrow\nu_{\rho}}(k;t=0)&=&\delta_{\rho\sigma};
\end{eqnarray}
\item due to the unitarity of $Z^{1/2}$,
\begin{eqnarray}\label{BC2}
\sum_{\rho}P^{(ret)}_{\nu_{\sigma}\rightarrow\nu_{\rho}}(k;t)&=&1;
\end{eqnarray}
\item from (\ref{Welement}),
\begin{eqnarray}\label{BC3}
P^{(ret)}_{\nu_{\sigma}\rightarrow\nu_{\rho}}(k;t)&=&P^{(ret)}_{\nu_{\rho}\rightarrow\nu_{\sigma}}(k;t);
\end{eqnarray}
\item 
under the condition of real $Z^{1/2}$(i.e. due to (\ref{a8}))
\begin{eqnarray}
 P^{(ret)}_{\nu_{\sigma}\rightarrow\nu_{\rho}}(k;t)=\left(|W_{\sigma\rho}(k;t)|^{2}+|W_{\sigma\bar\rho}(k;t)|^{2}\right)= P^{>}_{\nu_{\sigma}\rightarrow\nu_{\rho}}(k;t).
\end{eqnarray}
\end{enumerate}
\subsection{The case of $i{\cal S}^{(ret)}_{\rho\sigma}(\vec x,t)$}
The $\vec k$-components of $i{\cal S}^{(ret)}_{\rho\sigma}(\vec x-\vec y,t)$ defined as
\begin{eqnarray}
is^{(ret)}_{\rho\sigma}(k;t)\equiv \frac{1}{V}\int d\vec x\int d\vec y i{\cal S}^{(ret)}_{\rho\sigma}(\vec x,t;\vec y,0)e^{-i\vec k\vec x}e^{i\vec k\vec y}
\end{eqnarray}
become
\begin{eqnarray}
 is^{(ret)}_{\rho\sigma}(k;t)=\theta(t)\sum_{j,r}z_{\rho j}^{1/2}z_{\sigma j}^{1/2\ast}\left(\left\{\alpha_{j}(kr;t),\alpha^{\dagger}_{j}(kr;0)\right\}u_{j}(kr){\bar u}_{j}(kr)+\left\{\beta_{j}^{\dagger}(-kr;t),\beta_{j}(-kr;0)\right\}v_{j}(-kr){\bar v}_{j}(-kr)\right).
\end{eqnarray}
With this quantity we can define
\begin{eqnarray}
\Pi^{(ret)}_{\nu_{\sigma}\rightarrow\nu_{\rho}}(k;t)\equiv\frac{1}{4}Tr[s^{(ret)}_{\rho\sigma}(k;t)s^{(ret)\dagger}_{\rho\sigma}(k;t)],
\end{eqnarray}
and call it the probability due to the same reasoning as before.
For $t\geq 0$, we obtain
\begin{eqnarray}\label{Pi}
\Pi^{(ret)}_{\nu_{\sigma}\rightarrow\nu_{\rho}}(k;t)&\equiv&\frac{1}{4}\sum_{j,i,r}z_{\rho j}^{1/2}z_{\sigma j}^{1/2\ast}z_{\rho i}^{1/2\ast}z_{\sigma i}^{1/2}Tr[(u_{j}(kr)u_{j}^{\dagger}(kr)e^{-i\omega_{j}t}+v_{j}(-kr)v_{j}^{\dagger}(-kr)e^{i\omega_{j}t})\nonumber\\
&\mbox{\hspace*{10mm}}&(u_{i}(kr)u_{i}^{\dagger}(kr)e^{i\omega_{i}t}+v_{i}(-kr)v_{i}^{\dagger}(-kr)e^{-i\omega_{i}t})]\nonumber\\
&=&\frac{1}{2}\sum_{j,i}z_{\rho j}^{1/2}z_{\sigma j}^{1/2\ast}z_{\rho i}^{1/2\ast}z_{\sigma i}^{1/2}[\rho_{ij}^{2}(k)(e^{-i(\omega_{j}-\omega_{i})t}+c.c.)+\lambda_{ij}^{2}(k)(e^{-i(\omega_{j}+\omega_{i})t}+c.c.)]\nonumber\\
&=&\delta_{\rho\sigma}-2\sum_{j,i}z_{\rho j}^{1/2}z_{\sigma j}^{1/2\ast}z_{\rho i}^{1/2\ast}z_{\sigma i}^{1/2}[\rho_{ij}^{2}(k)sin^{2}(\frac{\omega_{j}-\omega_{i}}{2}t)+\lambda_{ij}^{2}(k)sin^{2}(\frac{\omega_{j}+\omega_{i}}{2}t)]\nonumber \\
&=&P^{(ret)}_{\nu_{\sigma}\rightarrow\nu_{\rho}}(k;t),
\end{eqnarray}
and this is equal to Eq.(\ref{BVprob}) in the two flavor case.
Furthermore we have the properties of this quantity
\begin{flushleft}
\begin{eqnarray}
(1)&\mbox{\hspace*{20mm}}&
\Pi^{(ret)}_{\nu_{\sigma}\rightarrow\nu_{\rho}}(k;t=0)=\delta_{\rho\sigma},\mbox{\hspace*{100mm}}\\
(2)&\mbox{\hspace*{20mm}}&
\sum_{\rho}\Pi^{(ret)}_{\nu_{\sigma}\rightarrow\nu_{\rho}}(k;t)=1,\\
(3)&\mbox{\hspace*{20mm}}&
\Pi^{(ret)}_{\nu_{\sigma}\rightarrow\nu_{\rho}}(k;t)=\Pi^{(ret)}_{\nu_{\rho}\rightarrow\nu_{\sigma}}(k;t),
\end{eqnarray}
\end{flushleft}
due to essentially the same reasons mentioned for deriving Eqs. (\ref{BC1}),(\ref{BC2}) and (\ref{BC3}).
As expected, 
\begin{flushleft}
\begin{eqnarray}
\mbox{\hspace*{4mm}}(4)&\mbox{\hspace*{20mm}}&
\Pi^{(ret)}_{\nu_{\sigma}\rightarrow\nu_{\rho}}(k;t)\rightarrow P^{>}_{\nu_{\sigma}\rightarrow\nu_{\rho}}(k;t)\mbox{ for the real $Z^{1/2}$,
\hspace*{100mm}}
\end{eqnarray}
\end{flushleft}
given explicitly by (\ref{a12}).
\section{DISCUSSION AND FINAL REMARKS}
We have generalized BV's formulae (\ref{BVprob}), $P_{\nu_{e}\rightarrow \nu_{\rho}}(k;t)$, to $P^{>}_{\nu_{\sigma}\rightarrow \nu_{\rho}}(k;t)$ of the many flavor case;
due to the unitarity of $Z^{1/2}$, these $P^{>}_{\nu_{\sigma}\rightarrow \nu_{\rho}}(k;t)$ satisfy the boundary conditions which are required for the probability interpretation, but are $\{\mu_{\lambda}\}$-dependent generally.
With those formulae, it has been shown that they are $\{\mu_{\lambda}\}$-independent and equal to the expectation values of the number operators for the real $Z^{1/2}$, which is of course the case of BV, i.e. the two flavor case.
At the same time, we have shown that the corresponding quantities, $\Pi^{>}_{\nu_{\sigma}\rightarrow \nu_{\rho}}(k;t)$ constructed on ${\cal H}_{m}$, cannot satisfy the bounday conditions.

On the other hand, we could construct the other quantities $P^{(ret)}_{\nu_{\sigma}\rightarrow \nu_{\rho}}(k;t)$ by employing the anti-commutators $\{\nu_{\rho}(\vec x,x^{0}), \nu^{\dagger}_{\sigma}(\vec y,y^{0})\}$;
these $P^{(ret)}_{\nu_{\sigma}\rightarrow \nu_{\rho}}(k;t)$ satisfy the boundary conditions stated above, due to the unitarity of $Z^{1/2}$.
They are automatically $\{\mu_{\lambda}\}$-independent because of the $\{\mu_{\lambda}\}$-independent c-number property of the anti-commutators, and are equal to the corresponding quantities $\Pi^{(ret)}_{\nu_{\sigma}\rightarrow \nu_{\rho}}(k;t)$ constructed on ${\cal H}_{m}$. 

Those quantities $P^{(ret)}_{\nu_{\sigma}\rightarrow \nu_{\rho}}(k;t)$ reduce to $P^{>}_{\nu_{\sigma}\rightarrow \nu_{\rho}}(k;t)$ for the real $Z^{1/2}$ and to BV's formulae $P_{\nu_{e}\rightarrow \nu_{\rho}}(k;t)$ for the two flavor case.
Then we conclude that the $\{\mu_{\lambda}\}$-independence of the "oscillation formulae" asserted by BV\cite{BV2} is essentially based on the $<$Setup$>$, and that the new factors in the BV's formulae, which are different from the usual oscillation formulae, are not the trace of the flavor Hilbert space construction, but come from the field theoretical treatment of mixing fields using Bogolyubov transformation.
In the present case, the coefficients of the new factors are fixed by the spin-$\frac{1}{2}$ property of neutrino.

The interrelationship among the relevant quantities is summarized in Figure 1.
\begin{center}
\begin{figure}[h]
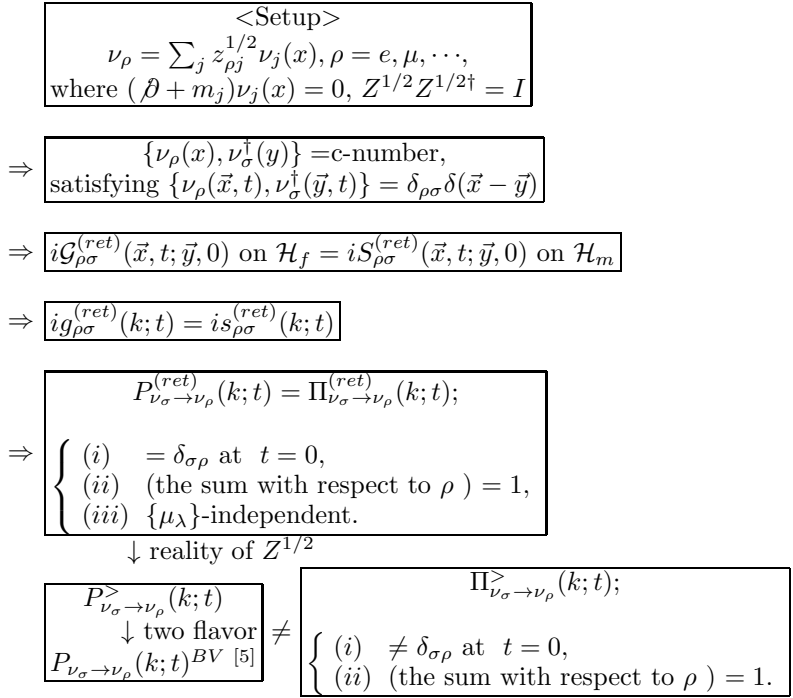

\begin{tabular}{clccc}
&
\begin{tabular}{|c|}
\hline
$<$Setup$>$\\
 $\nu_{\rho}=\sum_{j}z^{1/2}_{\rho j}\nu_{j}(x), \rho=e,\mu,\cdots$,\\
 where $(\not\partial+ m_{j})\nu_{j}(x) = 0$, $Z^{1/2}Z^{1/2\dagger}=I$\\
\hline
\end{tabular}
&&&\\
&&&&\\
$\Rightarrow$&
\begin{tabular}{|c|}
\hline
$\{\nu_{\rho}(x),\nu_{\sigma}^{\dagger}(y)\}=$c-number,\\
 satisfying $\{\nu_{\rho}(\vec x,t),\nu_{\sigma}^{\dagger}(\vec y, t)\}=\delta_{\rho\sigma}\delta(\vec x-\vec y)$
\\ \hline
\end{tabular}
&&&\\
&&&&\\
$\Rightarrow$&
\begin{tabular}{|c|}
\hline
$i{\cal G}^{(ret)}_{\rho\sigma}(\vec x,t;\vec y, 0) \mbox{ on }{\cal H}_{f} = iS^{(ret)}_{\rho\sigma}(\vec x,t;\vec y, 0) \mbox{ on } {\cal H}_{m} $
\\ \hline
\end{tabular}
&&&\\
&&&&\\
$\Rightarrow$&
\begin{tabular}{|c|}
\hline
$ig^{(ret)}_{\rho\sigma}(k;t)= is^{(ret)}_{\rho\sigma}(k;t)$
\\ \hline
\end{tabular}
&&&\\
&&&&\\
$\Rightarrow$&
\begin{tabular}{|c|}
\hline
$P^{(ret)}_{\nu_{\sigma}\rightarrow \nu_{\rho}}(k;t) = \Pi^{(ret)}_{\nu_{\sigma}\rightarrow \nu_{\rho}}(k;t);$\\
\\
$\left\{\begin{array}{ll}
(i)& = \delta_{\sigma\rho} \mbox{\  at \ } t=0,\\
(ii)& \mbox{(the sum with respect to $\rho$ ) $ =1$},\\
(iii)& \{\mu_{\lambda}\}\mbox{-independent}.
\end{array}\right.$
\\ \hline
\end{tabular}
&&&\\
 &  
{$\mbox{\hspace*{10mm}}
\downarrow $ reality of $Z^{1/2}$}&&& \\
&
\begin{tabular}{|c|}
\hline
$P^{>}_{\nu_{\sigma}\rightarrow \nu_{\rho}}(k;t)$\\
\hfill$\downarrow$
 two flavor\\
$P_{\nu_{\sigma}\rightarrow \nu_{\rho}}(k;t)^{BV\cite{BV2}}$\\
\hline
\end{tabular}
%
$\neq$
\begin{tabular}{|c|}
\hline
$ \Pi^{>}_{\nu_{\sigma}\rightarrow \nu_{\rho}}(k;t);$\\
\\
$\left\{\begin{array}{ll}
(i)& \neq \delta_{\sigma\rho} \mbox{\  at \ } t=0,\\
(ii)& \mbox{(the sum with respect to $\rho$ ) $ =1$}.
\end{array}\right.$\\
\hline
\end{tabular}
&&&
\end{tabular}
\newline
\newline
\caption{Without reality of $Z^{1/2}$, $P^{>}_{\nu_{\sigma}\rightarrow \nu_{\rho}}(k;t) \neq \langle N_{\sigma};kr;t\rangle  _{\rho-f}$, and $P^{>}_{\nu_{\sigma}\rightarrow \nu_{\rho}}(k;t)$ is $\{\mu_{\lambda}\}$-dependent for $N_{f}\geq 3$.}
\end{figure}
\end{center}

Let us give some remarks as follows.
Concerning the construction of ${\cal H}_{f}$, we give in Appendix C a remark that we cannot necessarily eliminate the possibility to construct ${\cal H}_{f}$ within the extent of the paper by Giunti et al.\cite{GKL}.
According to the context of the present paper, the construction of ${\cal H}_{f}$ is not always excluded, since there are some quantities, such as $\langle 0(0) |  \{\nu_{\rho}(x), \bar \nu_{\sigma} (y)\}| 0(0)\rangle$, which are obtained on the basis of ${\cal H}_{m}$ and equal to ones obtained on the basis of ${\cal H}_{f}$.
For the quantities constructed on ${\cal H}_{f}$, $\{\mu_{\lambda}\}$-independence seems to suggest that there is no difference of those quantities from the ones constructed on ${\cal H}_{m}$.
It is a future task to make clearer the field theoretical basis of such an anticipation.

In the present paper, we examined only the contributions from the propagator by extracting a part of neutrino propagation from the full transition amplitudes corresponding to various neutrino experiments.
The relation between the oscillation probability discussed above and the full transition probability is not clear yet, and then we cannot decide which propagator one should use to calculate the oscillation probability.
It may be a meaningful fact that $\Pi^{(ret)}_{\nu_{\sigma}\rightarrow \nu_{\rho}}(k;t)$ satisfies the boundary conditions while $\Pi^{>}_{\nu_{\sigma}\rightarrow \nu_{\rho}}(k;t)$ does not.

Here it seems worthy to remark on the treatment of low-energy weak processes  with accompanying neutrinos, such as $\pi^{+}\rightarrow l^{+}_{\rho}\nu_{\rho}$, $\mu^{+}\rightarrow e^{+}\nu_{e}\bar\nu_{\mu}$ and $n\rightarrow p e\nu_{e}$.
In the lowest-order calculation with respect to the weak interaction, each participating flavor neutrino has been treated as an asymptotic field with a definite mass nearly equal to zero, and we have obtained important informations on the structure of the weak interactions.
As an illustration, we examine the decay probabilities of $\pi^{+}\rightarrow l^{+}_{\rho}\nu_{\rho}$, $\rho=\mu, e$.
By employing the usual V-A weak interaction, we have the decay probabilities for $\pi^{+}\rightarrow l^{+}_{\rho}\nu_{\rho}$,
\begin{eqnarray}\label{p-mu}
     P(\pi^{+}\rightarrow l^{+}_{\rho}\nu_{\rho})=&&\frac{f^{2}_{\pi}G^{2}_{\beta}}{8\pi}m^{2}_{\rho}\left(1+\frac{\mu^{2}_{\rho}}{m^{2}_{\rho}}-\frac{(m^{2}_{\rho}-\mu^{2}_{\rho})^{2}}{m^{2}_{\pi}m^{2}_{\rho}}\right)m_{\pi}\nonumber\\
&&\times \left((1-\frac{(m_{\rho}+\mu_{\rho})^{2}}{m^{2}_{\pi}})(1-\frac{(m_{\rho}-\mu_{\rho})^{2}}{m^{2}_{\pi}})\right)^{1/2};
\end{eqnarray}
here, $f_{\pi}$ is the pion decay constant defined by $\langle 0|A_{\alpha}(0)|\pi^{+}(p)\rangle =if_{\pi}p_{\alpha}$, $A_{\alpha}(x)=$ the weak axial current such as $i\bar d(x)\gamma_{\alpha}\gamma_{5}u(x)$, $(f_{\pi})_{exp}\simeq 131MeV$;
$G_{\beta}\simeq (10^{-5}/M^{2}_{p})$ is the weak Fermi coupling constant(including Cabbibo-Kobayashi-Maskawa angle);
$m_{\rho}$ and $\mu_{\rho}$ are the masses of $l_{\rho}$ and $\nu_{\rho}$ respectively.
For $\mu_{\rho}=0$, (\ref{p-mu}) reduces to 
\begin{eqnarray}\label{p}
     P(\pi^{+}\rightarrow l^{+}_{\rho}\nu_{\rho})=&&\frac{f^{2}_{\pi}G^{2}_{\beta}}{8\pi}m^{2}_{\rho}m_{\pi}
\left(1-\frac{m^{2}_{\rho}}{m^{2}_{\pi}}\right)^{2};
\end{eqnarray}
then we obtain numerical values which are in good agreement with the experimental ones.

The calculation above is based on the existence of asymptotic flavor neutrino field with definite mass.
But we can not regard the flavor neutrino fields $\nu_{\rho}(x)$'s to define the asymptotic fields, leading to time-independent creation- and annihilation-operators with definite 4-momenta, since in accordance with the $<$Setup$>$ with mass differences among $m_{j}$'s, each $\alpha_{\rho}(kr;t)$ and $\beta^{\dagger}_{\rho}(-kr;t)$ in the expansion of $\nu_{\rho}(x)$ dose not have a simple time dependence, since $\alpha_{\rho}(kr;t)=\sum_{j}\{{\cal K}(k)_{\rho j}\alpha_{j}(kr;t)+{\cal K}(k)_{\rho \bar j}\beta^{\dagger}_{j}(-kr;t)\}$.
Contrary, each of $\alpha_{j}(kr;t)$ and $\beta^{\dagger}_{j}(-kr;t)$ in the expansion of $\nu_{j}(x)$ has a simple time dependence as $exp(\mp i\omega_{j}(k)t)$ with $\omega_{j}(k)=(\vec k^{2}+m^{2}_{j})^{1/2}$;
thus we can construct the Hilbert space ${\cal H}_{m}$, independently of the time, by employing the operators $\alpha^{\dagger}_{j}(kr)\equiv\alpha^{\dagger}_{j}(kr;t)e^{-i\omega_{j}(k)t}$ and $\beta^{\dagger}_{j}(-kr)\equiv\beta^{\dagger}_{j}(-kr;t)e^{-i\omega_{j}(k)t}$.
The probabilities $P(\pi^{+}\rightarrow l^{+}_{\rho}\nu_{\rho})$ are given by
\begin{eqnarray}\label{p-m}
     P(\pi^{+}\rightarrow l^{+}_{\rho}\nu_{\rho})=&&\frac{f^{2}_{\pi}G^{2}_{\beta}}{8\pi}m^{2}_{\rho}m_{\pi}\sum_{j}{}'|z^{1/2}_{\rho j}|^{2}
\left(1+\frac{m^{2}_{j}}{m^{2}_{\rho}}-\frac{(m^{2}_{\rho}-m^{2}_{j})^{2}}{m^{2}_{\pi}m^{2}_{\rho}}\right)\nonumber\\
&&\times \left\{(1-\frac{(m_{\rho}+m_{j})^{2}}{m^{2}_{\pi}})(1-\frac{(m_{\rho}-m_{j})^{2}}{m^{2}_{\pi}})\right\}^{1/2};
\end{eqnarray}
where the summation $\sum_{j}'$ is performed over $j$'s which are allowed under 4-momentum conservation.
We see that (\ref{p-m}) reduce to (\ref{p}) when 
\begin{enumerate}
\item 
masses of the relevant neutrinos are negligibly small, and even if a neutrino has a nonnegligible mass, the corresponding mixing angle, $|z^{1/2}_{\rho j}|$ is very small, and
\item
$\sum_{j}'|z^{1/2}_{\rho j}|^{2}=1$ holds almost precisely.
\end{enumerate}
Though (\ref{p-mu}) and (\ref{p-m}) give the same formula (\ref{p}) in the limit of massless neutrinos, one should adopt the well-defined formula, (\ref{p-m}). It is necessary for us to examine whether all experimental data are consistent or not with those conditions as noted above.
\newline

The oscillation formulae usually employed are possibly modified due to various reasons, e.g. due to  certain effects coming from a more detailed quantum-mechanical or field theoretical description\cite{FT} reflecting real experimental situations
\footnote{Recently, along this line, a possible approach has been done by one of the present author(T.Y.) and Ishikawa\cite{YI}.}
, or due to the mixing of the known left-handed neutrinos with some right-handed neutrinos which may propagate in the bulk space including some extra dimensions\cite{ED}.
Anyhow, it may be necessary to investigate the neutrino oscillation by applying the field theory in accordance with respective experimental situations.

\begin{center}
                   {\bf ACKNOWLEDGMENTS}
\end{center}
 
    The authors would like to express their thanks to Prof. K.Ishikawa and the other members of the particle theory group in Hokkaido University for their helpful discussions. 
Thanks are also due to Prof. G.Vitiello and Dr. M.Blasone, Salerno University, for their communications and remarks.

\appendix
\section{$\mu_{\lambda}$-independence of 
$P^{>}_{\nu_{\sigma}\rightarrow \nu_{\rho}}$ under CP-invariance condition}
%
As an example to see the $\mu_{\lambda}$-independence of (\ref{P>}), we can derive the equality
\begin{eqnarray}\label{a1}
&&|\{\alpha_{\rho}(kr;t),\alpha_{\sigma}^{\dagger}(kr;0)\}|^{2}+|\{\beta_{\rho}^{\dagger}(kr;t),\alpha_{\sigma}^{\dagger}(kr;0)\}|^{2}\nonumber\\
&&=|\{\alpha_{\rho}^{BV}(kr;t),\alpha_{\sigma}^{BV\dagger}(kr;0)\}|^{2}+|\{\beta_{\rho}^{BV}(kr;t),\alpha_{\sigma}^{BV\dagger}(kr;0)\}|^{2}
\end{eqnarray}
under the reality condition of $Z$.
Here $\alpha_{\rho}^{BV}$ and $\beta_{\rho}^{BV}$ are the operators employed in Ref.\cite{FHY};
\begin{eqnarray}
\left(\begin{array}{c}
{\bf \alpha}_{F}(kr;t)\\{\bf \beta}_{F}^{\dagger}(-kr;t)
\end{array}\right)
=\left(\begin{array}{cc}
{\bf \rho}_{F}(k)&i{\bf \lambda}_{F}(k)\\i{\bf \lambda}_{F}(k)&{\bf \rho}_{F}(k)
\end{array}\right)
\left(\begin{array}{c}
{\bf \alpha}^{BV}_{F}(kr;t)\\{\bf \beta}_{F}^{BV\dagger}(-kr;t)
\end{array}\right)
\end{eqnarray}
where ${\bf \rho}_{F}(k)$ and ${\bf \lambda}_{F}(k)$ are $N_{f}\times N_{f}$ diagonal matrices;
\begin{eqnarray}
{\bf \rho}_{F}(k)=\left(\begin{array}{ccc}
\rho_{e1}(k)&&0\\&\rho_{\mu 2}(k)&\\0&&\ddots
\end{array}\right),
{\bf \lambda}_{F}(k)=\left(\begin{array}{ccc}
\lambda_{e1}(k)&&0\\&\lambda_{\mu 2}(k)&\\0&&\ddots
\end{array}\right)
\end{eqnarray}
For simplicity, we use the shortcut notations as
\begin{eqnarray}
\begin{array}{c}
\alpha_{\rho}(kr;t)\longrightarrow \alpha_{\rho}(t),\\
\beta_{\rho}(kr;t)\longrightarrow \beta_{\rho}(t),
\end{array}
\end{eqnarray}
then
\begin{eqnarray}\label{a5}
\{\alpha_{\rho}(t),\alpha_{\sigma}^{\dagger}(0)\}&=&\{\sum_{j}(z^{1/2}_{\rho j}\rho_{\rho j}\alpha_{j}(t)+iz^{1/2}_{\rho j}\lambda_{\rho j}\beta_{j}^{\dagger}(t)),\sum_{l}(z^{1/2\ast}_{\sigma l}\rho_{\sigma l}\alpha_{l}^{\dagger}(0)-iz^{1/2\ast}_{\sigma l}\lambda_{\sigma l}\beta_{l}(0))\}\nonumber\\
&=&\sum_{j}z^{1/2}_{\rho j}z^{1/2\ast}_{\sigma j}(\rho_{\rho j}\rho_{\sigma j}e^{-i\omega_{j}t}+\lambda_{\rho j}\lambda_{\sigma j}e^{i\omega_{j}t})=W_{\sigma\rho}(k;t)^{\ast},
\end{eqnarray}
and
\begin{eqnarray}\label{a6}
\{\beta_{\rho}^{\dagger}(t),\beta_{\sigma}(0)\}
&=&\sum_{j}z^{1/2}_{\rho j}z^{1/2\ast}_{\sigma j}(\lambda_{\rho j}\lambda_{\sigma j}e^{-i\omega_{j}t}+\rho_{\rho j}\rho_{\sigma j}e^{i\omega_{j}t})=W_{\bar\sigma\bar\rho}(k;t)^{\ast},\nonumber\\
\{\beta_{\rho}^{\dagger}(t),\alpha_{\sigma}^{\dagger}(0)\}
&=&\sum_{j}z^{1/2}_{\rho j}z^{1/2\ast}_{\sigma j}i(\lambda_{\rho j}\rho_{\sigma j}e^{-i\omega_{j}t}-\rho_{\rho j}\lambda_{\sigma j}e^{i\omega_{j}t})=W_{\sigma\bar\rho}(k;t)^{\ast},\nonumber\\
\{\alpha_{\rho}(t),\beta_{\sigma}(0)\}
&=&\sum_{j}z^{1/2}_{\rho j}z^{1/2\ast}_{\sigma j}i(-\rho_{\rho j}\lambda_{\sigma j}e^{-i\omega_{j}t}+\lambda_{\rho j}\rho_{\sigma j}e^{i\omega_{j}t})=W_{\bar\sigma\rho}(k;t)^{\ast}.
\end{eqnarray}
For the real $Z^{1/2}$, there are some relations among these anti-commutators as
\begin{eqnarray}\label{a2}
\{\beta_{\rho}^{\dagger}(t),\beta_{\sigma}(0)\}=\{\alpha_{\rho}^{\dagger}(t),\alpha_{\sigma}(0)\},\mbox{\hspace*{1cm}}
\{\alpha_{\rho}(t),\beta_{\sigma}(0)\}=-\{\beta_{\rho}(t),\alpha_{\sigma}(0)\},
\end{eqnarray}
i.e.
\begin{eqnarray}\label{a8}
W_{\bar\sigma\bar\rho}(k;t)^{\ast}=W_{\sigma\rho}(k;t),\mbox{\hspace*{1cm}}W_{\bar\sigma\rho}(k;t)^{\ast}=-W_{\sigma\bar\rho}(k;t).
\end{eqnarray}
(These relations (\ref{a8}) are also confirmed directly from (\ref{W}) by noting $P^{\dagger}=(z^{1/2}_{\rho j}\rho_{\rho j})^{\dagger}=P^{T}$ and $\Lambda^{\dagger}=(z^{1/2}_{\rho j}\lambda_{\rho j})^{\dagger}=\Lambda^{T}$ for real $Z^{1/2}$.)
By taking $(e,1)$, $(\mu,2), (\tau,3)\cdots$ for $(\rho,j)$ and $(\sigma,l)$,
\begin{eqnarray}\label{a3}
\mbox{L.H.S. of (\ref{a1})}=&&|\{\rho_{\rho j}\alpha^{BV}_{\rho}(t)+i\lambda_{\rho j}\beta^{BV\dagger}_{\rho}(t),\rho_{\sigma l}\alpha^{BV\dagger}_{\sigma}(0)-i\lambda_{\sigma l}\beta^{BV}_{\sigma}(0)\}|^{2}\nonumber\\
&&+|\{i\lambda_{\rho j}\alpha^{BV}_{\rho}(t)+\rho_{\rho j}\beta^{BV\dagger}_{\rho}(t),\rho_{\sigma l}\alpha^{BV\dagger}_{\sigma}(0)-i\lambda_{\sigma l}\beta^{BV}_{\sigma}(0)\}|^{2}\nonumber\\
=&&\rho_{\sigma l}^{2}|\{\alpha_{\rho}^{BV}(t),\alpha_{\sigma}^{BV\dagger}(0)\}|^{2}+\lambda_{\sigma l}^{2}|\{\beta_{\rho}^{BV\dagger}(t),\beta_{\sigma}^{BV}(0)\}|^{2}\nonumber\\
&&+\lambda_{\sigma l}^{2}|\{\alpha_{\rho}^{BV}(t),\beta_{\sigma}^{BV}(0)\}|^{2}+\rho_{\sigma l}^{2}|\{\beta_{\rho}^{BV\dagger}(t),\alpha_{\sigma}^{BV\dagger}(0)\}|^{2}\nonumber\\
&&+i\rho_{\sigma l}\lambda_{\sigma l}\{\alpha_{\rho}^{BV}(t),\alpha_{\sigma}^{BV\dagger}(0)\}\{\alpha_{\rho}^{BV\dagger}(t),\beta_{\sigma}^{BV\dagger}(0)\}+H.c.\nonumber\\
&&+i\rho_{\sigma l}\lambda_{\sigma l}\{\beta_{\rho}^{BV\dagger}(t),\alpha_{\sigma}^{BV\dagger}(0)\}\{\beta_{\rho}^{BV}(t),\beta_{\sigma}^{BV\dagger}(0)\}+H.c..
\end{eqnarray}
Employing (\ref{a2}),
\begin{eqnarray}
\mbox{(\ref{a3})}=|\{\alpha_{\rho}^{BV}(kr;t),\alpha_{\sigma}^{BV\dagger}(kr;0)\}|^{2}+|\{\beta_{\rho}^{BV\dagger}(kr;t),\alpha_{\sigma}^{BV\dagger}(kr;0)\}|^{2},
\end{eqnarray}
leading to (\ref{a1}).
It is needless to note that $P^{(ret)}_{\nu_{\sigma}\rightarrow \nu_{\rho}}(k;t)$ given by (\ref{Pret}) is shown to be equal to that calculated for the special choice of $\mu_{\lambda}$'s as above without requiring the reality of $Z^{1/2}$.

On the other hand, we obtain
\begin{eqnarray}\label{a4}
\mbox{the last side of (\ref{P>})}&=&
\sum_{i,j}[{\cal K}_{\sigma j}\phi_{j}{\cal K}^{\ast}_{\rho j}+{\cal K}_{\sigma \bar j}\phi^{\ast}_{j}{\cal K}^{\ast}_{\rho \bar j}]
[{\cal K}_{\rho i}\phi^{\ast}_{i}{\cal K}^{\ast}_{\sigma i}+{\cal K}_{\rho \bar i}\phi_{i}{\cal K}^{\ast}_{\sigma \bar i}]\nonumber\\
&&+\sum_{i,j}[{\cal K}_{\sigma j}\phi_{j}{\cal K}^{\ast}_{\bar\rho j}+{\cal K}_{\sigma \bar j}\phi^{\ast}_{j}{\cal K}^{\ast}_{\bar\rho \bar j}]
[{\cal K}_{\bar\rho i}\phi^{\ast}_{i}{\cal K}^{\ast}_{\sigma i}+{\cal K}_{\bar\rho \bar i}\phi_{i}{\cal K}^{\ast}_{\sigma \bar i}]\nonumber\\
&=&\sum_{i,j}z^{1/2}_{\sigma j}z^{1/2\ast}_{\rho j}z^{1/2}_{\rho i}z^{1/2\ast}_{\sigma i}
[\rho_{\sigma j}\rho_{\sigma i}\rho_{ji}\phi_{j}\phi_{i}^{\ast}
+\lambda_{\sigma j}\lambda_{\sigma i}\rho_{ji}\phi_{j}^{\ast}\phi_{i}\nonumber\\
&&-\rho_{\sigma j}\lambda_{\sigma i}\lambda_{ji}\phi_{j}\phi_{i}
+\lambda_{\sigma j}\rho_{\sigma i}\lambda_{ji}\phi_{j}^{\ast}\phi_{i}^{\ast}].
\end{eqnarray}
For the case of real $Z^{1/2}$,
\begin{eqnarray}\label{a12}
\mbox{(\ref{a4})}&=&\sum_{i,j}z^{1/2}_{\sigma j}z^{1/2}_{\rho j}z^{1/2}_{\rho i}z^{1/2}_{\sigma i}[\rho_{ji}^{2}(\phi_{j}\phi_{i}^{\ast}+\phi_{j}^{\ast}\phi_{i})+\lambda_{ji}^{2}(\phi_{j}\phi_{i}+\phi_{j}^{\ast}\phi_{i}^{\ast})]/2\nonumber\\
&=&\delta_{\sigma\rho}-2\sum_{i,j}z^{1/2}_{\sigma j}z^{1/2}_{\rho j}z^{1/2}_{\rho i}z^{1/2}_{\sigma i}[\rho_{ji}^{2}sin^{2}(\frac{\Delta\omega_{ji}}{2}t)+\lambda_{ji}^{2}sin^{2}(\frac{\omega_{j}+\omega_{i}}{2}t)],
\end{eqnarray}
and then we get a $\mu_{\lambda}$-independent quantity explicitly.

\section{Number operator}
Let us define the expectation values of the number operator of neutrino;
\begin{eqnarray}
\langle N_{\sigma};kr;t\rangle  _{\rho-f}\equiv {}_{f}\langle 0(t)|\alpha_{\rho}(kr;t)N_{\sigma}(t=0)\alpha_{\rho}^{\dagger}(kr;t)|0(t)\rangle  _{f},
\end{eqnarray}
where
\begin{eqnarray}
N_{\sigma}(t)&\equiv&\sum_{\vec q,s}(n_{\sigma}(qs;t)-\bar n_{\sigma}(-qs;t)),\nonumber\\
n_{\sigma}(qs;t)&\equiv&\alpha_{\sigma}^{\dagger}(qs;t)\alpha_{\sigma}(qs;t),\nonumber\\
\bar n_{\sigma}(-qs;t)&\equiv&\beta_{\sigma}^{\dagger}(-qs;t)\beta_{\sigma}(-qs;t),\nonumber\\
\langle n_{\sigma}(qs;0);kr;t\rangle  _{\rho-f}&\equiv& {}_{f}\langle 0(t)|\alpha_{\rho}(kr;t)n_{\sigma}(qs;0))\alpha_{\rho}^{\dagger}(kr;t)|0(t)\rangle  _{f},\nonumber\\
\langle \bar n_{\sigma}(-qs;0);kr;t\rangle  _{\rho-f}&\equiv& {}_{f}\langle 0(t)|\alpha_{\rho}(kr;t)\bar n_{\sigma}(-qs;0))\alpha_{\rho}^{\dagger}(kr;t)|0(t)\rangle  _{f}.
\end{eqnarray}
>From (\ref{DefOfW})
\begin{eqnarray}
n_{\sigma}(qs;0)=&&\sum_{\lambda,\kappa}\{W_{\sigma\lambda}(q;t)^{\ast}\alpha_{\lambda}^{\dagger}(qs;t)+W_{\sigma\bar\lambda}(q;t)^{\ast}\beta_{\lambda}(-qs;t)\}
\{W_{\sigma\kappa}(q;t)\alpha_{\kappa}(qs;t)+W_{\sigma\bar\kappa}(q;t)\beta_{\kappa}^{\dagger}(-qs;t)\},\nonumber\\
\langle n_{\sigma}(qs;0)&&;kr;t\rangle  _{\rho-f}=|W_{\sigma\rho}(k;t)|^{2}\delta_{rs}\delta(\vec k,\vec q)+\sum_{\lambda}|W_{\sigma\bar\lambda}(q;t)|^{2},\\
\bar n_{\sigma}(-qs;0)=&&\sum_{\lambda,\kappa}\{W_{\bar\sigma\lambda}(q;t)\alpha_{\lambda}(qs;t)+W_{\bar\sigma\bar\lambda}(q;t)\beta_{\lambda}^{\dagger}(-qs;t)\}
\{W_{\bar\sigma\kappa}(q;t)^{\ast}\alpha_{\kappa}^{\dagger}(qs;t)+W_{\bar\sigma\bar\kappa}(q;t)^{\ast}\beta_{\kappa}(-qs;t)\},\nonumber\\
\langle \bar n_{\sigma}(-qs;0)&&;kr;t\rangle  _{\rho-f}=\sum_{\lambda}|W_{\bar\sigma\lambda}(q;t)|^{2}-|W_{\bar\sigma\rho}(k;t)|^{2}\delta_{rs}\delta(\vec k,\vec q).
\end{eqnarray}
We have
\begin{eqnarray}
\langle N_{\sigma};kr;t\rangle  _{\rho-f}&=&|W_{\sigma\rho}(k;t)|^{2}+|W_{\bar\sigma\rho}(k;t)|^{2}+2\sum_{\vec q,\lambda}\left(|W_{\sigma\bar\lambda}(q;t)|^{2}-|W_{\bar\sigma\lambda}(q;t)|^{2}\right);
\end{eqnarray}
then for $t\geq 0$
\begin{eqnarray}
\langle N_{\sigma};kr;t\rangle  _{\rho-f}=P^{>}_{\nu_{\sigma}\rightarrow\nu_{\rho}}(k;t)+|W_{\bar\sigma\rho}(k;t)|^{2}-|W_{\sigma\bar\rho}(k;t)|^{2}+2\sum_{\vec q,\lambda}\left(|W_{\sigma\bar\lambda}(q;t)|^{2}-|W_{\bar\sigma\lambda}(q;t)|^{2}\right),
\end{eqnarray}
from which we obtain under the reality condition of $Z^{1/2}$(or (\ref{a8}))
\begin{eqnarray}
\langle N_{\sigma};kr;t\rangle  _{\rho-f}&=&P^{>}_{\nu_{\sigma}\rightarrow\nu_{\rho}}(k;t),\\
\sum_{\sigma}\langle N_{\sigma};kr;t\rangle  _{\rho-f}&=&1.
\end{eqnarray}

Thus we see that, in order to derive the oscillation formulas for $N_{f}$-flavors as the generalization of those given by B.V.\cite{BV2}, we have to assume the reality of $Z^{1/2}$, supported by $T$-(or $CP$-)invariance\cite{KOW}.

\section{On ${\cal H}_{f}$ construction}
%
We give a remark on the construction of the flavor Hilbert space ${\cal H}_{f}$ in connection with the work of Giunti et al.\cite{GKL}.
In Ref.\cite{GKL}, the Majorana neutrino field is considered for simplicity, but we consider the Dirac one in accordance with the representations as explained in Sec.II.

First of all, we consider the expansion of the flavor-neutrino field $\nu_{\rho}(x)$ as Eq.(\ref{n}),
\begin{eqnarray}\label{nu1}
\nu_{\rho}(x)
&=&\sum_{j=1}^{N_{f}}z^{1/2}_{\rho j}\nu_{j}(x)\nonumber\\
&=&\frac{1}{\sqrt{V}}\sum_{j=1}^{N_{f}}z^{1/2}_{\rho j}\sum_{\vec k r}e^{i\vec k\cdot\vec x}\{u_{j}(kr)\alpha_{j}(kr;t)+v_{j}(-kr)\beta^{\dagger}_{j}(-kr;t)\},
\end{eqnarray}
where $\rho=e,\mu,\tau,\cdots$.
With the 2-component spinors of spin eigenstates $w(k\uparrow)$, the momentum-helicity eigensolutions are given as
\begin{eqnarray}
u_{j}(k\uparrow)=\left(\begin{array}{c}c_{j}\\ s_{j} \end{array}\right)\otimes w(k\uparrow),\mbox{\hspace*{5mm}} 
u_{j}(k\downarrow)=\left(\begin{array}{c}s_{j}\\ c_{j} \end{array}\right)\otimes w(k\downarrow),\nonumber\\
v_{j}(k\uparrow)=\left(\begin{array}{c}-s_{j}\\ c_{j} \end{array}\right)\otimes w(k\downarrow),\mbox{\hspace*{5mm}} 
v_{j}(k\downarrow)=\left(\begin{array}{c}c_{j}\\ -s_{j} \end{array}\right)\otimes w(k\uparrow),
\end{eqnarray}
where
\begin{eqnarray}
w(k\uparrow)\equiv \left(\begin{array}{c}a(\vec k)\\ b(\vec k) \end{array}\right),\mbox{\hspace*{5mm}}
w(k\downarrow)\equiv \left(\begin{array}{c} -b^{\ast}(\vec k)\\ a^{\ast}(\vec k) \end{array}\right);\\
\frac{\vec\sigma \vec k}{|\vec k|}w(k,r)=\left\{\begin{array}{c}+1\\ -1 \end{array}\right\}w(k,r) \mbox{ for }=\left\{\begin{array}{c}\uparrow\\ \downarrow \end{array}\right\};
\end{eqnarray}
$a(\vec k)=cos({\vartheta}/2)e^{-i{\varphi}/2}$, $b(\vec k)=sin({\vartheta}/2)e^{i{\varphi}/2}$ for $k_{z}=kcos({\vartheta})$, $k_{x}+ik_{y}=ksin({\vartheta}) e^{i{\varphi}}$;
$c_{j}=cos(\chi_{j}/2)$, $s_{j}=sin(\chi_{j}/2)$ with $cot(\chi_{j})=|\vec k|/m_{j}$.
By noting $a(-\vec k)=-ib^{\ast}(\vec k)$ and  $b(-\vec k)=ia^{\ast}(\vec k)$, we have\begin{eqnarray}
v_{j}(-k\uparrow)=\left(\begin{array}{c}-s_{j}\\ c_{j} \end{array}\right)\otimes iw(k\uparrow), \mbox{\hspace*{5mm}}
v_{j}(-k\downarrow)=\left(\begin{array}{c}c_{j}\\ -s_{j} \end{array}\right)\otimes iw(k\downarrow).
\end{eqnarray}
Following the notation employed in Ref.\cite{GKL}, $c_{j}$ and $s_{j}$ are written as $\kappa_{j+}$ and $\kappa_{j-}$ respctively;
\begin{eqnarray}
\kappa_{j\pm}=\sqrt{\frac{\omega_{j}(k)\pm |\vec k|}{2\omega_{j}(k)}}.
\end{eqnarray}
Thus (\ref{nu1}) is expressed as
\begin{eqnarray}\label{nu2}
\nu_{\rho}(x)
=&&\frac{1}{\sqrt{V}}\sum_{j}z^{1/2}_{\rho j}\sum_{\vec k}e^{i\vec k\cdot\vec x}\left[\alpha_{j}(k\uparrow ;t)\left(\begin{array}{c}\kappa_{j+}\\ \kappa_{j-} \end{array}\right)\otimes w(k\uparrow)+\alpha_{j}(k\downarrow ;t)\left(\begin{array}{c}\kappa_{j-}\\ \kappa_{j+} \end{array}\right)\otimes w(k\downarrow)\right.\nonumber\\
&&\left.+i\beta_{j}^{\dagger}(-k\uparrow ;t)\left(\begin{array}{c}-\kappa_{j-}\\ \kappa_{j+} \end{array}\right)\otimes w(k\uparrow)+i\beta^{\dagger}_{j}(-k\downarrow ;t)\left(\begin{array}{c}\kappa_{j+}\\ -\kappa_{j-} \end{array}\right)\otimes w(k\downarrow)\right].
\end{eqnarray}
By defining further
\begin{eqnarray}\label{DefOfAB}
A_{\rho\pm}(kr;t)\equiv\sum_{j}z^{1/2}_{\rho j}\alpha_{j}(kr ;t)\kappa_{j\pm},\mbox{\hspace*{5mm}}
B_{\rho\pm}^{\dagger}(-kr;t)\equiv\pm i\sum_{j}z^{1/2}_{\rho j}\beta_{j}^{\dagger}(-kr ;t)\kappa_{j\pm},
\end{eqnarray}
(\ref{nu2}) is rewritten as
\begin{eqnarray}\label{nu3}
\nu_{\rho}(x)
=&&\frac{1}{\sqrt{V}}\sum_{\vec k}e^{i\vec k\cdot\vec x}\left[
\left(\begin{array}{c}A_{\rho +}(k\uparrow ;t)\\ A_{\rho -}(k\uparrow ;t) \end{array}\right)\otimes w(k\uparrow)
+\left(\begin{array}{c}A_{\rho -}(k\downarrow ;t)\\ A_{\rho +}(k\downarrow ;t) \end{array}\right)\otimes w(k\downarrow)\right.\nonumber\\
&&\left.+\left(\begin{array}{c}B_{\rho -}^{\dagger}(-k\uparrow ;t)\\ B_{\rho +}^{\dagger}(-k\uparrow ;t) \end{array}\right)\otimes w(k\uparrow)
+\left(\begin{array}{c}B_{\rho +}^{\dagger}(-k\downarrow ;t)\\ B_{\rho -}^{\dagger}(-k\downarrow ;t) \end{array}\right)\otimes w(k\downarrow)
\right].
\end{eqnarray}
>From the canonical commutation relations at the equal time among $\alpha_{j},\alpha_{j}^{\dagger},\beta_{j}$ and $\beta_{j}^{\dagger}$, we obtain
\begin{eqnarray}\label{commutator}
&&\{A_{\sigma\pm}(kr;t),A_{\rho\pm}^{\dagger}(qs;t)\}=\{B_{\rho\pm}(kr;t),B_{\sigma\pm}^{\dagger}(qs;t)\}=\sum_{j}z^{1/2}_{\sigma j}z^{1/2\ast}_{\rho j}\frac{\omega_{j}(k)\pm |\vec k|}{2\omega_{j}(k)}\delta_{rs}\delta(\vec k,\vec q),\nonumber\\
&&\{A_{\sigma\pm}(kr;t),A_{\rho\mp}^{\dagger}(qs;t)\}=\{B_{\rho\pm}(kr;t),B_{\sigma\mp}^{\dagger}(qs;t)\}=\sum_{j}z^{1/2}_{\sigma j}z^{1/2\ast}_{\rho j}\frac{|m_{j}|}{2\omega_{j}(k)}\delta_{rs}\delta(\vec k,\vec q),\nonumber\\
&&\mbox{others}=0,
\end{eqnarray}
but these are not the canonical commutation relations at the equal time among $A_{\sigma\pm},A_{\sigma\pm}^{\dagger},B_{\sigma\pm}$ and $B_{\sigma\pm}^{\dagger}$, and we cannot construct the flavor Hilbert space ${\cal H}_{f}$ with these operators.

In the extremely relativistic limit, (\ref{commutator}) reduces to
\begin{eqnarray}
&&\{A_{\sigma +}(kr;t),A_{\rho +}^{\dagger}(qs;t)\}\longrightarrow \delta_{\sigma\rho}\delta_{rs}\delta(\vec k,\vec q),\nonumber\\
&&\{B_{\sigma +}(kr;t),B_{\rho +}^{\dagger}(qs;t)\}\longrightarrow \delta_{\sigma\rho}\delta_{rs}\delta(\vec k,\vec q),\nonumber\\
&&\mbox{others}=0.
\end{eqnarray}
Then the high-momentum part of R.H.S of (\ref{nu3}) becomes
\begin{eqnarray}
\nu_{\rho}(x)
\longrightarrow
\frac{1}{\sqrt{V}}\sum_{\vec k}e^{i\vec k\cdot\vec x}\{
\left(\begin{array}{c}A_{\rho +}(k\uparrow ;t)\\ B_{\rho +}^{\dagger}(-k\uparrow ;t) \end{array}\right)\otimes w(k\uparrow)
+\left(\begin{array}{c}B_{\rho +}^{\dagger}(-k\downarrow ;t)\\ A_{\rho +}(k\downarrow ;t) \end{array}\right)\otimes w(k\downarrow)\}
\end{eqnarray}
with
\begin{eqnarray}
A_{\rho +}(kr;t)=\sum_{j}z^{1/2}_{\rho j}\alpha_{j}(kr ;t), \mbox{\hspace*{5mm}}
B_{\rho +}^{\dagger}(-kr;t)=\sum_{j}z^{1/2}_{\rho j}\beta_{j}^{\dagger}(-kr ;t).
\end{eqnarray}
If we follow the assertion in Ref.\cite{GKL}, we can construct the Hilbert space ${\cal H}_{f}$ only in the extremely relativistic limit.
This assertion, however, is self-evident, since in this limit the mass differences among neutrinos do not play any physical role.
Further, the transformation (\ref{DefOfAB}) is not canonical, and the number of operators $(A_{\rho \pm}(kr;t),B_{\rho \pm}^{\dagger}(-kr;t))$ is twice larger than that of $(\alpha_{j}(kr;t),\beta_{j}^{\dagger}(-kr;t))$.
Thus, the assertion against the ${\cal H}_{f}$ construction seems not appropriate.

Here we show an example which has the canonical commutation relations and the possibility of constructing the flavor Hilbert space ${\cal H}_{f}$.

We define a kind of Bogolyubov transformation by
\begin{eqnarray}\label{neu_expantion}
\tilde A_{\rho}(kr;t)&\equiv&\sum_{j}z^{1/2}_{\rho j}[\alpha_{j}(kr;t)\kappa_{j+}-i\beta_{j}^{\dagger}(-kr;t)\kappa_{j-}],\nonumber\\
\tilde B_{\rho}^{\dagger}(-kr;t)&\equiv&\sum_{j}z^{1/2}_{\rho j}[-i\alpha_{j}(kr;t)\kappa_{j-}+\beta_{j}^{\dagger}(-kr;t)\kappa_{j+}].
\end{eqnarray}
We have the canonical communication relations as
\begin{eqnarray}
\{\tilde A_{\rho}(kr;t),\tilde A_{\rho}^{\dagger}(qs;t)\}&=&\delta_{rs}\delta(\vec k,\vec q)=\{\tilde B_{\rho}(-kr;t),\tilde B_{\rho}^{\dagger}(-qs;t)\},\nonumber\\
others&=&0,
\end{eqnarray}
and in this case the neutrino field is expanded as
\begin{eqnarray}\label{nu4}
\nu_{\rho}(x)
=&&\frac{1}{\sqrt{V}}\sum_{\vec k}e^{i\vec k\cdot\vec x}\left[
\left(\tilde A_{\rho}(k\uparrow ;t)\left(\begin{array}{c}1\\ 0 \end{array}\right)+i\tilde B_{\rho}^{\dagger}(-k\uparrow ;t)\left(\begin{array}{c}0\\ 1 \end{array}\right)\right)\otimes w(k\uparrow)\right.\nonumber\\
&&\left.+\left(\tilde A_{\rho}(k\downarrow ;t)\left(\begin{array}{c}0\\ 1 \end{array}\right)+i\tilde B_{\rho}^{\dagger}(-k\downarrow ;t)\left(\begin{array}{c}1\\ 0 \end{array}\right)\right)\otimes w(k\downarrow)\right],
\end{eqnarray}
which corresponds to the plane-wave expansion with $\mu_{\lambda}=0$.
Then physical quantities, which should be $\{\mu_{\lambda}\}$-independent, are allowed to be calculated by employing the expansion (\ref{nu4}) with (\ref{neu_expantion});
the quantities such as $\rho_{\sigma j}(k)=cos((\chi_{\sigma}-\chi_{j})/2)$ and $\lambda_{\sigma j}(k)=sin((\chi_{\sigma}-\chi_{j})/2)$ are now replaced by $cos(\chi_{j}/2)=c_{j}$ and $sin(-\chi_{j}/2)=-s_{j}$;
thus the matrices $[P(k)_{\rho j}]$ and $[\Lambda(k)_{\rho j}]$ appearing in ${\cal K}(k)$ are replaced by $[z^{1/2}_{\rho j}\rho_{j}(k)]$ and $[-z^{1/2}_{\rho j}\lambda_{j}(k)]$.


\end{document}